\documentclass[modern,trackchanges]{aastex62}
\usepackage{amssymb, amsmath}
\usepackage{morefloats}
\usepackage{hyperref}
\usepackage{pifont}
\usepackage[T1]{fontenc}

\NewPageAfterKeywords
\begin{document}

\title{Shedding Light on the Isolation of Luminous Blue Variables}

\author{Erin Aadland}
\affiliation{Lowell Observatory, 1400 W Mars Hill Road, Flagstaff, AZ 86001, USA}
\affiliation{Department of Physics and Astronomy, Northern Arizona University, Flagstaff, AZ, 86011-6010, USA}

\author{Philip Massey}
\affiliation{Lowell Observatory, 1400 W Mars Hill Road, Flagstaff, AZ 86001, USA}
\affiliation{Department of Physics and Astronomy, Northern Arizona University, Flagstaff, AZ, 86011-6010, USA}

\author{Kathryn F. Neugent}
\affiliation{Department of Astronomy, University of Washington, Seattle, WA, 98195, USA}
\affiliation{Lowell Observatory, 1400 W Mars Hill Road, Flagstaff, AZ 86001, USA}

\author{Maria R. Drout}
\affiliation{Department of Astronomy and Astrophysics, University of Toronto, 50 St.\ George Street, Toronto, Ontario, M5S 3H4 Canada}
\affiliation{The Observatories of the Carnegie Institution for Science, 813 Santa Barbara St., Pasadena, CA 91101, USA}

\begin{abstract}
    In the standard view of massive star evolution, luminous blue variables (LBVs) are transitional objects between the most massive O-type stars and Wolf-Rayet (WR) stars.  With short lifetimes, these stars should all be found near one another. A recent study of LBVs in the Large Magellanic Cloud (LMC) found instead that LBVs are considerably more isolated than either O-type stars or WRs, with a distribution intermediate between that of the WRs and red supergiants (RSGs). A similar study, using a more restricted sample of LBVs, reached the opposite conclusion. Both studies relied upon the distance to the nearest spectroscopically identified O-type star to define the degree of isolation.  However, our knowledge of the spectroscopic content of the LMC is quite spotty.  Here we re-examine the issue using carefully defined photometric criteria to select the highest-mass unevolved stars (``bright blue stars,'' or BBSs), using spatially complete photometric catalogs of the LMC, M31, and M33.  Our study finds that the LBVs are no more isolated than BBSs or WRs. This result holds no matter which sample of LBVs we employ.  A statistical test shows that we can rule out the LBVs having the same distribution as the RSGs, which are about $2\times$ more isolated.  We demonstrate the robustness of our results using the second-closest neighbor.   Furthermore, the majority of LBVs in the LMC are found in or near OB associations as are the BBS and WRs;  the RSGs are not. We conclude that the spatial distribution of LBVs therefore is consistent with the standard picture of massive star evolution.

\end{abstract}
\keywords{stars: evolution -- stars: massive -- stars: early-type -- supergiants -- stars: Wolf-Rayet}

\section{Introduction}
\label{Sec-Intro}

	Luminous blue variables (LBVs) are currently at the center of an evolution controversy.  Traditionally, LBVs are thought to be a short-lived transitional stage in the lives of the most massive stars (>50$M_\odot$), subsequently evolving to a Wolf-Rayet (WR) star (see, e.g. discussion in \citealt{Meynet2011, Massey2013}). As a high-mass star evolves off the hydrogen burning main sequence, these stars reach their atmospheric Eddington limit, undergoing episodic  mass-loss rates on the order of $10^{-4}M_\odot$ yr$^{-1}$ while becoming many magnitudes brighter (see, e.g. \citealt{Humphreys94,LamersBook}).  Some of the most well-known examples of LBVs are S Dor in the Large Magellanic Cloud (LMC),  P Cyg and $\eta$ Car in the Milky Way, and the Hubble-Sandage variables \citep{Hubble1953} in M31 and M33.  
	
Observationally, this picture has been supported by the LBV-outbursts of R127 and HDE~269582 in the LMC (\citealt{Walborn2017} and references therein).  These stars were once classified as WN9/Ofpe (``slash'') stars, with properties intermediate between O-type stars and low-excitation (late-type) WN-type WRs.  

Nevertheless, this view has never been universally accepted, and recent discoveries by the supernova/transient community have further clouded this picture.  Examples include the discovery of a class of supernova (``Type IIn'') whose progenitors undergo eruptions (ejecting 0.01 $-$ 1 $M_\odot$ of material) in the years \emph{immediately} preceding core-collapse \cite[][and references therein]{Smith2014} plus the identification in nearby galaxies of relatively low-luminosity transients (``supernova impostors''), whose outbursts resemble those of LBVs in some ways, but are occasionally linked to stars of only 8$-$10 $M_\odot$  (e.g., \citealt{Thompson2009}).  In addition, recent work has shown that the locations of Type IIn SN and some lower luminosity transients do \emph{not} correlate with tracers of recent star formation (e.g. H$\alpha$ and UV emission) in their host galaxies \cite[e.g.,][]{Habergham2014}. Together, these observations indicate that ``LBV-like'' eruptive behavior may not always represent a transitional phase and may not always be linked to the highest-mass stars, but this is to be expected given the wide range of instabilities that may occur in stars.

    Another argument that has been raised against the standard picture of LBVs is the question of their isolation.  For instance, \citet{Gallagher81} argued that the Hubble-Sandage variable AF And was in an inter-arm region of M31, and hence not in an area of recent star formation.  However, \citet{Madore78} made the point about the distribution of supergiants in M33,  that massive stars are occasionally found in isolation, emphasizing that each generation of astronomers seems to rediscover this anew, and indeed \citet{MC83}, \citet{MasseyMCIMF}, and \citet{NeugentM33} confirmed spectroscopically the presence of O-type and WR stars in relatively isolated regions of the Magellanic Clouds, M31, and M33.  
    
More recently, the question of the isolation of LBVs has been investigated statistically, rather than antidotally, by \cite{Smith2015}.  They found that LBVs do not share the same spatial distribution as their alleged massive star progenitors.  In the standard picture, LBVs should be a very short-lived stage in the lives of the most massive stars, and thus should be found in the same locales.  \citet{Smith2015} measured the projected distance from each LBV to the closest spectroscopically known O-type star using the SIMBAD database, finding instead that LBVs are considerably more isolated than WRs (which they are expected to evolve into) or O-type stars (which they are expected to have evolved from).  Instead, their isolation is intermediate between that of the WRs and red supergiants (RSGs), which \cite{Smith2015} interpret as meaning that the lower limit to LBV initial masses  are 12-15$M_\odot$, much lower than in the standard picture where LBVs are descended from stars of much higher masses\footnote{Although the most luminous RSGs come from stars with initial masses of 30$M_\odot$, most RSGs are expected to be descended from stars of 15$M_\odot$ and below when one takes into account both the IMF-weighted numbers and the time spent in the RSG phase.}.  Based on these results, they propose that LBVs are the mass gainers in binary systems, as then the LBV progenitors would be of lower mass than in the standard picture, allowing them to have longer lifetimes and disperse from their birthplace. (See \citealt{model} for a model matching these results.) The LBV may also receive a kick by its companion's SN explosion, resulting in further isolation.
    
    \cite{Humphreys2016} re-examined the isolation question using a more conservative definition of what actually constitutes an LBV, and employed the same technique as \cite{Smith2015} by using the SIMBAD database to identify the closest known O-type star.  \cite{Humphreys2016} subdivided their sample up into ``classical'' and ``less luminous'' LBVs and found that the isolation of ``classical'' LBVs was statistically indistinguishable from O-type stars, while the ``less luminous'' sample had a distribution more similar to RSGs.  They also included a qualitative discussion of the fact that most of the Hubble-Sandage variables were found in regions rich with supergiants in M31 and M33, consistent with their connection to high-mass stars. \cite{Smith2016} debated whether subdividing the LBVs as \cite{Humphreys2016} did was statistically justified, and countered that even if the LBVs are split into these subcategories the conclusions of \cite{Smith2015} remained valid.

	However, there are several inherent problems with the technique of selecting the nearest spectroscopically identified O-type star in the LMC.  As \cite{Neugent2018} points out, our knowledge of the spectroscopic content of the Magellanic Clouds is quite spotty, with more spectroscopic studies restricted to a few well-studied regions. New spectroscopic studies of any star-forming region in the LMC typically uncover numerous previously unidentified O-type stars (see, e.g., \citealt{Massey2000}).  Furthermore, most O-type stars will be late-type O-types dwarfs with masses as low as 15-20$M_\odot$ (see Table 1 of \citealt{Massey2017}), not the high-mass stars that are expected to evolve to WRs via an LBV phase.  Indeed, \cite{Smith2018} went on to use the same technique to examine the isolation of a new class of WR stars in the LMC (WN3/O3 stars; see \citealt{Neugent2017}) and concluded they were also ``extremely isolated.'' Based on these results, they proposed that the WN3/O3 stars are lower-mass helium stars in binaries that have been stripped of their hydrogen envelopes (e.g. \citealt{Goth2018}).  \cite{Neugent2018} re-examined this conclusion by identifying the nearest unevolved massive stars using the LMC photometry catalog of \citet{Zaritsky}. (The \citealt{Zaritsky} catalog covers all of the LMC uniformly and goes many magnitudes deeper than the O-type star population; thus, incompleteness will not be an issue except in the most crowded regions.)  When \cite{Neugent2018} did that, they found that the WN3/O3s had the same distribution as the single early-type WNs in the LMC.  Their result does not eliminate WN3/O3s possible origin as stripped binaries, but it no longer requires it.  Rather, their distribution is consistent with them having a common origin with other early WNs.

	We therefore thought it would be worthwhile to weigh in on the LBV spatial distribution in the LMC using the \cite{Zaritsky} photometry catalog rather than relying upon the incomplete spectroscopic knowledge of the O-type star population, especially since most O-type stars are not expected to be the progenitors of LBVs in any event.  In addition, we also apply this photometric technique to the LBV population of M31 and M33 using the Local Group Galaxy Survey (LGGS) photometry \citep{Massey2007_II,LGGS_sources} to determine how robust our results are, and whether different trends are seen in different galaxies at different metallicities. 
    
    We will first give an overview of our selection of the various samples in Section~\ref{Sec-samples}.  In Section~\ref{Sec-LMC}, we examine the projection angular separation of the LBVs in relation to the other star samples.  Next, we evaluate the isolation of LBVs in M31 and M33 with their sample selection in Section~\ref{Sec-M}.  We will end the paper by reviewing our results and the implications for the origins of LBVs in Section~\ref{Sec-conclusions}. \\

\section{Sample Selection and Rationale}
 \label{Sec-samples}
 
In this section, we will describe in some detail the rationale for the sample selection, concentrating on the LMC, and how these will be modified for M31 and M33. 

\subsection{Unevolved Massive Stars}
	Selecting unevolved massive stars photometrically is complicated.  There are two problems, both related to their high temperatures.  First, the optical colors of unevolved massive stars change very little over the relevant effective temperature range of 30000 K (a B0 star) to 50000 (the hottest O-type stars), and yet the bolometric corrections change very significantly ($\sim$1.5 mag) over this range (see, e.g., \citealt{Massey1989}).  Thus, from optical photometry alone, one cannot infer effective temperatures \textit{or} bolometric luminosities to determine masses from evolutionary tracks with any precision\footnote{This problem can be partially relieved by using spacecraft UV photometry, but is then sensitive to variations in the shape of the UV extinction law;  see discussion in \citet{MasseyIMF}.}.  Second, as a  massive star ages, it cools slightly, with the absolute visual magnitude getting much brighter as the flux distribution shifts to longer wavelengths.  Thus during its main-sequence evolution, a 25$M_\odot$ star will evolve from an O6 V star with $M_V=-3.8$ to a B0 I with $M_V=-5.4$ (see Table 1 in \citealt{Massey2017}).  Although we would like to define photometric criteria for the identification of unevolved massive stars as a $V$ limit as a function of color (or, preferably, a reddening-free index), the relative insensitivity of optical colors to effective temperature (and the lack of reliable synthetic colors from model atmospheres) renders this impractical given finite errors in the photometry.  

	Instead, we have adopted a broad-brush approach by using the photometry of \cite{Zaritsky} of the LMC to select high-mass stars, following \cite{Neugent2018}.  Our goal is to select the highest-mass stars. We adopt a magnitude cut-off of $V<13.9$ to restrict the sample to stars with $M_V<-5$ for an apparent distance modulus of 18.9, based upon a distance of 50 kpc \citep{vandenBergh} and a typical reddening of the OB stars of $E(B-V)=0.13$ \citep{MasseyMCIMF, Massey2007_II}.  We construct the standard reddening-free index $Q$ as $(U-B)-0.72\times(B-V)$, and then restrict the sample to the hottest stars, $Q<-0.88$, where $Q<-0.88$ corresponds roughly to an effective temperature $>$ 35000 K \citep{MasseyMCIMF}. In order to exclude stars with unrealistic colors (indicating bad photometry), we further restrict the sample to $Q>-1.2$, $(U-B)<-0.5$, and $B-V<0.2$.   For M31 and M33, we will use the LGGS photometry \citep{Massey2007_II,LGGS_sources}, adopting an appropriate $V$ cut-off to achieve the same $M_V<-5$ criterion, and using the same color requirements.

\begin{figure}
\epsscale{1}
\plotone{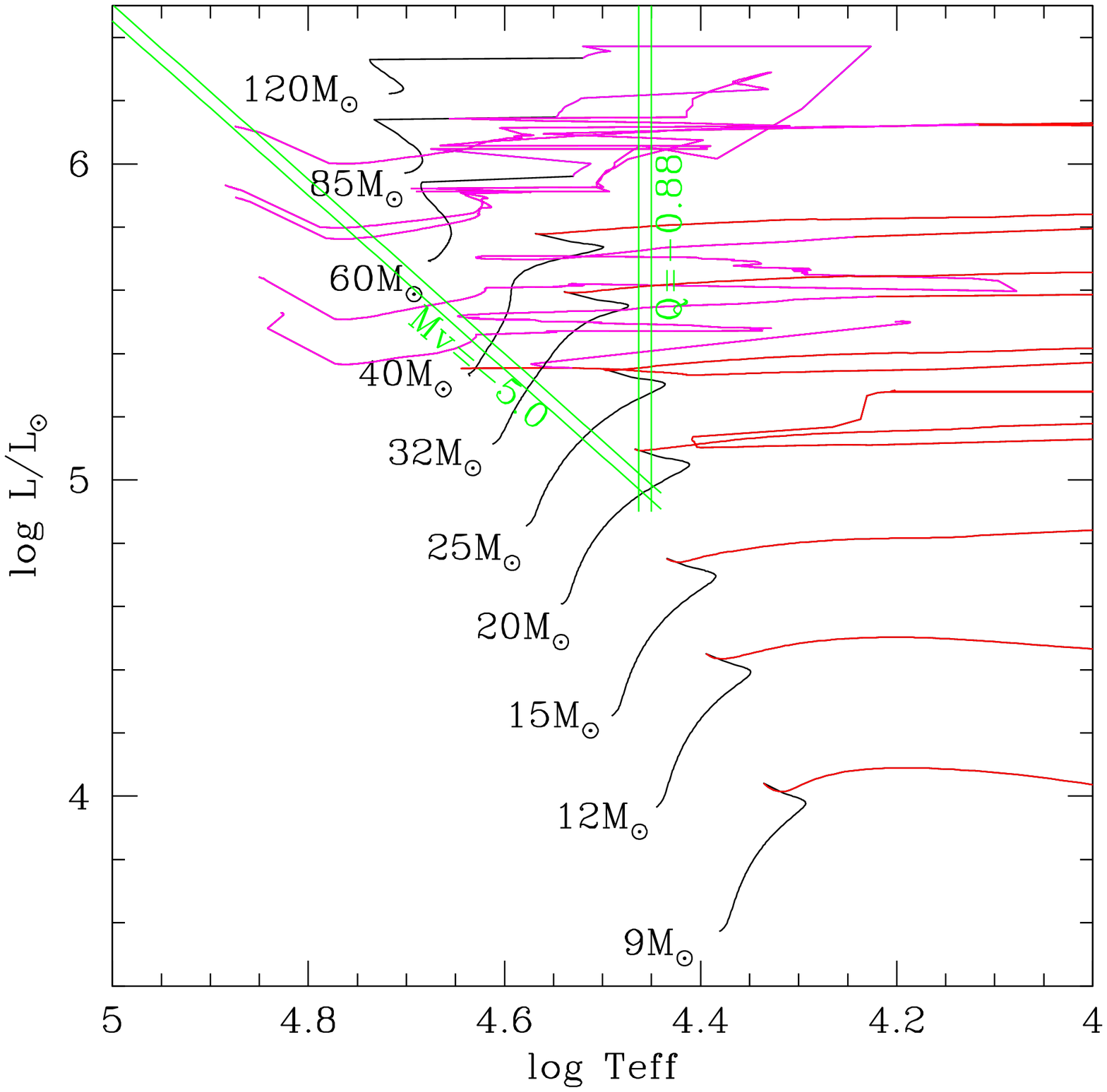}
\caption{\label{fig:mass} BBS selection.  The region between the two double sets of green lines indicate the location in the HRD of the BBS sample. Weighted by the lifetimes within this region, and by the IMF, we expect that most of these unevolved stars will have masses $>40M_\odot$. The evolution tracks are from \citet{Sylvia}, with black denoting the main-sequence phase, and the colored regions indicating later stages; magenta denotes the WR stage.  The two pairs of diagonal lines are based upon the $M_V=-5.0$ cutoff, with the bolometric correction computed using relations in \cite{MasseyPuls05} [upper] and \cite{Martins06} [lower].  The two vertical lines denote the $Q=-0.88$ cutoff using color transformations derived using the \cite{Martins06} colors [left] and the ATLAS9 \citep{Kurucz} colors tabulated on the Castelli web site http://wwwuser.oats.inaf.it/castelli/colors.html [right].}
\end{figure}

	  As shown in Figure~\ref{fig:mass}, these samples should be relatively complete for stars of 40$M_\odot$ and above (except in crowded regions), but will include stars as low in mass as 25$M_\odot$ near the end of core-H burning when they are at their brightest visually, but still within our narrow range of $Q$.  We designate this sample as ``bright blue stars'' (BBSs); we expect them all to be high-mass stars, primarily of O-type (with some high-mass early B-supergiants), although certainly not all O-type stars will be in our sample, as stars of 15$M_\odot$ are late-type O stars early in their lives (see Table 1 in \citealt{Massey2017}) and are too faint and not intrinsically blue enough to be included.  These lower-mass O-type stars are not expected to have an LBV phase by standard single-star evolutionary theory, but should instead evolve into RSGs.  Therefore, by selecting only the highest-mass stars, we are looking at those that evolve into LBVs according to the standard model.

\subsection{Luminous Blue Variables}
	\cite{Smith2015}, \cite{Humphreys2016}, and \cite{LBVlist} all disagree on exactly which stars should be considered LBVs and LBV candidates.  This has been an issue since the very beginning of the terminology of ``luminous blue variable'' (see \citealt{Bohannan1997}), and remains controversial today. The stars that were originally used by \cite{Conti1984} to define the class included S Dor, P Cyg, $\eta$ Car, and the Hubble-Sandage variables.  However, the ``sanctification'' of LBV candidates (luminous blue stars which are spectroscopically similar to confirmed LBVs but which have not shown spectacular photometric outbursts) is contentious (see, e.g., discussion in \citealt{Massey2007,Clark2012,Humphreys2014,Humphreys2017}) and there is even disagreement on what constitutes a valid LBV candidate with some stars being dismissed as merely ``warm hypergiants'' \citep{Humphreys2017}.
    
    This disagreement resulted in \cite{Humphreys2016} removing six stars that \cite{Smith2015} had included as being LBVs, leaving a total of 10 stars that both \cite{Smith2015} and \cite{Humphreys2016} agreed upon.  The \cite{LBVlist} catalog includes these 10 agreed-upon LBVs and 17 additional LBVs and LBV candidates.  Stars in \cite{LBVlist} were included as candidate LBVs if they had been described as having spectra that were similar to established LBVs; see, for example, \cite{Massey2007}.  The LBVs from each source are listed in Table \ref{tab:LMC}.

\begin{figure}
\epsscale{1}
\plotone{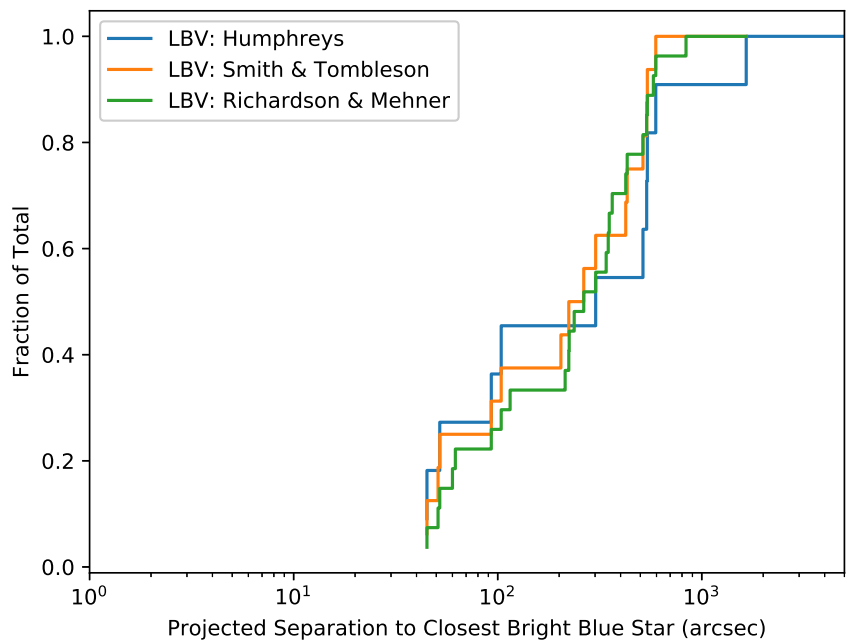}
\caption{\label{fig:list} Comparison of the projected separations for three LBV samples in the LMC.  The LBV sets from \cite{Smith2015}, \cite{Humphreys2016}, and \cite{LBVlist} compared to each other via projected separation to the nearest BBS.}
\end{figure}

	Since one of the disagreements between the isolation studies is over which LBVs should be included, we looked at the projected angular separation from their nearest BBS neighbor for each of the three LMC sets (\citealt{Smith2015}, \citealt{Humphreys2016}, and \citealt{LBVlist}) of LBVs, which is displayed in Figure \ref{fig:list}.  Visually, distributions of these LBV sets do not appear significantly different from each other. 
	
	In order to evaluate whether the sets can be considered as coming from the same parent distribution a Kolmogorov-Smirnov (KS) test (see, e.g., \citealt{Press}) was implemented.  The KS test outputs a p-value, which if less than 5\%, suggests that the two samples are from different parent distributions. Large values are consistent with the samples being from the same parent population, but of course this does not prove that they are.  We find that the p-value between the sets \cite{Humphreys2016} and \cite{Smith2015} is 82\%, between the sets \cite{LBVlist} and \cite{Smith2015} is 100\%, and between the sets \cite{Humphreys2016} and \cite{LBVlist} is 47\%.  Since all three KS tests have p-values that are well in excess of the 5\% criterion described above, there is no evidence of the three LBV samples having different distributions of projected offsets from the BBS sample.   Thus, since the \cite{LBVlist} catalog is the most up-to-date list of LBVs and LBV candidates, we have drawn our LBV samples from it.  We will use this same source for our choice of M31 and M33 LBV candidates. (Note that we will henceforth refer to these stars as ``LBVs,'' although many are only ``candidates,'' based upon their spectroscopic similarity to the more classical LBVs.)

\subsection{Evolved Massive Stars}
	We will also compare the separation of the LBVs with respect to WRs and RSGs.  One of the most intriguing results to come from the \cite{Smith2015} study was that the separations of LBVs from the nearest spectroscopically confirmed O-type star was significantly greater than that of the O-type stars themselves or that of WRs.  If confirmed, this would demonstrate that LBVs could not be a transitional phase between O-type stars and WRs. 

	All three galaxies (LMC, M31, and M33) have been subject to recent surveys of WRs which are believed to be complete at the 5\% or better level \citep{NeugentM33,NeugentM31,Neugent2018}. 
	
	For RSGs, spectroscopically identified samples are available for all three galaxies \citep{Drout2012,Neugent2012,MasseyEvans}, but we know that these samples are not complete, and so we used photometry to select our RSG samples.  For the LMC, we are using the same criteria as used successfully by \citet{Neugent2012}, using  the 2MASS catalog \citep{2MASS} with the cuts of $1.2 \geq (J-K) > 0.9$ and $K \leq 10.2$. For a typical RSG with with a spectral type of M1-1.5, the effective temperature will be 3700~K and have a bolometric correction at K of +2.8~mag \citep{LevesqueMC}. This corresponds to a bolometric luminosity of about $\log L/L_\odot=4.1$, or about 9$M_\odot$ and above. For M31 and M33, we use  $V<20$ and $V-R>0.6$ to select our photometric sample, and then use color cuts in {\it B-V} to remove foreground stars, following \citet{Massey98}.  The $V<20$ criteria corresponds to a $\log L/L_\odot=4.5$ when applying a bolometric correction at $V$ of $-1.5$ mag, or about 12$M_\odot$ and above.
	
\subsection{Removing Foreground Stars}
	Given our use of photometric selection, what do we expect by way of foreground contaminations in this study?   Low-luminosity Galactic stars can be confused with extragalactic supergiants of similar colors, i.e., nearby red dwarfs and extragalactic red supergiants. Our previous studies have used the Besan\k{c}on model of the Milky Way \citep{ZeFrench} to examine the extent of foreground contamination in the color-magnitude diagram in the direction of the Magellanic Clouds (Figure~1 in \citealt{Neugent2010}), M31 (Figure 1 in \citealt{DroutM31}), and M33 (Figure 2 in \citealt{Drout2012}). We expect little or no contamination from foreground objects for the BBSs, where the only low-luminosity stars hot and blue enough would be white dwarfs and sdOs (subdwarf O stars).   Using the photometric criteria defined above, the Besan\k{c}on model does not predict a single such contaminant in either the LMC sample of BBSs, or in the combined M31/M33 data set.  Our extensive spectroscopy in M31/M33 \citep{LGGS_sources} has actually found seven foreground white dwarfs (three of these meet the photometric criteria we are using), reinforcing the fact that the space density of white dwarfs is not particularly well known locally, although progress is being made in this area (see, e.g., \citealt{BianchiWDs}).   As for red stars, we have found that using proper motions was an effective tool to eliminate contamination of the LMC sample (see \citealt{Neugent2012}), while for M31 and M33 we have successfully employed two-color diagrams to separate foreground stars and RSGs \citep{Massey98,Drout2012,MasseyEvans}.   Here we use utilize Data Release 2 (DR2) {\it Gaia} parallaxes for the first time to eliminate foreground stars seen toward the LMC.  
	
	For stars in the LMC, we check for foreground contamination in our photometrically selected samples of BBSs and RSGs using proper motions ($\mu_{\alpha}$, $\mu_{\delta}$) and parallaxes ($\pi$) from the \emph{Gaia} DR2. To identify probable members of the LMC and likely foreground stars we follow a procedure based on that described in \citet{Gaia2018}. In brief: we first download {\it Gaia} sources within a 5$^\circ$ radius of the LMC center (5:15:20 -69:20:10). Restricting ourselves to sources with relative parallax error $\pi$/$\sigma_{\pi}$ $<$5 and magnitudes G $<$ 17.5 mag, we determined the median proper motions and median parallax for the sample. As in \citet{Gaia2018}, we then further eliminated any sources whose $\mu_{\alpha}$ or $\mu_{\delta}$ were more than four times the robust scatter estimate. Using this sample, we then determined the covariance matrix {\bf $\mu$} of $\mu_{\alpha}$, $\mu_{\delta}$, and $\pi$ and defined a filter on proper motion and parallax, requiring that the transpose of $\mu$ with $\sigma^{-1}$ times $\mu$ ({\bf $\mu^T \sigma^{-1} \mu$}) is greater than 12.8 (corresponding to the 99.5\% confidence region) for classification as a probable foreground star. We then applied this filter to the samples of the LMC BBSs and RSGs described above, after cross matching with the \emph{Gaia} database.  
	
For the M31 and M33 samples, DR2 does not go quite faint enough to be useful, and so we rely upon photometric means to separate the foreground contamination for the extragalactic sample.  For the M31/M33 BBS samples, we rely upon the expectation that there will be little or no contamination for very blue stars. \\

\section{Large Magellanic Cloud}
\label{Sec-LMC}

\subsection{Sample Selection}
\label{Sec-LMC-Sample}

Since we are relying upon ground-based photometry to select our BBS sample (using the \citealt{Zaritsky} catalog) and our RSG sample (using 2MASS), we decided to exclude stars from all of our samples which lay within 10 arcmin of the center of 30 Dor, as crowding is severe there, particularly in the core R136 region.  

{\bf BBS:} With that exclusion, there are 688 stars in our BBS.  After removing 22 {\it Gaia}-selected  foreground stars, our reference sample is left with 666 stars.  
{\bf LBVs:} The exclusion region around 30 Dor requires that we remove R143, leaving 26 LBVs and LBV candidates.  {\bf WRs:} The WR stars were selected based on ``The Fifth Catalog of LMC Wolf-Rayet Stars'' given in \cite{Neugent2018}, excluding R127 and HDE ~269582,  the two former ``slash'' stars that are currently in an LBV state \citep{Walborn2017}.  After excluding stars near 30~Dor, we have 117 WRs in our sample.   {\bf RSGs:} We selected candidate RSGs using the 2MASS list, as described above.   The initial list contained 1288 stars.  Our {\it Gaia} study then eliminated 105 (9\%) of these sources as likely foreground, leaving us with 1183\footnote{Note that the 1288 number is significantly smaller than the 1949 RSG candidates identified photometrically by \cite{Neugent2012}. In order to obtain the cleanest sample we could we only used 2MASS sources whose photometric quality flags were ``A.''  The actual number of LMC RSGs is likely 1500-1700; we will address this more exactly in a future paper. Of course, these numbers are much larger than the 505 LMC RSGs that \cite{Neugent2012} confirmed from radial velocity measurements, as only a subset of their photometric candidates were observed spectroscopically.}.  As described above, this sample will be dominated by stars with masses of 9-15$M_\odot$.

Our initial motivation for using the photometric catalog rather than spectroscopically identified O-type stars was that we were concerned about the spotty coverage of previous spectroscopic surveys in the LMC, where observers have typically concentrated on probing the stellar content of particular OB associations 
(see, e.g., \citealt{Massey117,Garmany58,Massey2000,MasseyMCIMF}).  Even in the era of wide-field multi-fiber surveys, such studies have concentrated on specific regions, e.g., the VLT-FLAMES Tarantula Survey of the 30~Dor region (\citealt{30Dor} and the subsequent papers in this series).  Furthermore, the SIMBAD database used by \cite{Smith2015} and \citet{Humphreys2016} are constantly evolving and incorporating new data, and there is no way of knowing which stars were included at the time of those studies.  We can investigate the spectroscopic completeness question here by using the current updated version of the \citet{Skiff} on-line catalog of spectral types, only some of which would have been incorporated into SIMBAD at the time of the \citep{Smith2015} study. 
    
	Of the 666 BBS (excluding foreground stars in the 30 Dor region), only 135 (20\%) have been identified spectroscopically as O-type stars. Another 140 (21\%) are luminous early B-type supergiants, and the remainder (59\%) have no spectral types. The presence of very early B-type supergiants in our sample is to be expected, considering stellar evolution and our use of a magnitude-limited sample, as early B-type supergiants can be more massive than many O-type stars are; only stars of 40$M_\odot$ and above are expected to bypass the early B supergiant phase in their evolution (again, see Table~1 in \citealt{Massey2017}). Conversely, a late-type O star may be as low in mass as 15$M_\odot$. However, the early B-type supergiants will be brighter visually as their bolometric corrections are not as significant: a 25$M_\odot$ B0~I star will be as bright visually as a 60$M_\odot$ O3~V star. This raises another problem with the methodology utilized in previous papers on the subject, as only the ``nearest O-type'' stars were considered, excluding B-supergiants that may have been more massive.  
	
	We note that with some irony that of the 22 BBS that were eliminated as ``probable foreground,'' about half had been spectroscopically confirmed as O-type stars, and therefore members of the LMC.  Possibly our method for eliminating stars based upon $\emph{Gaia}$ was a bit too aggressive, due, perhaps, to  errors in $\emph{Gaia}$ proper motions/parallaxes for stars in dense regions or multiple systems or unusual kinematics for a small set of massive LMC stars.  Alternatively, it could be that some prior spectral classifications mistook sdO stars for the real thing. Given the predictions of the Besan\k{c}on model and individual examination of the discrepant stars, we suspect the former.  We have run the samples both with and without these \textit{Gaia}-selected foreground stars included with indistinguishable results.

\subsection{Analysis}
\label{Sec-LMC-Analysis}

\begin{figure}
\epsscale{1}
\plotone{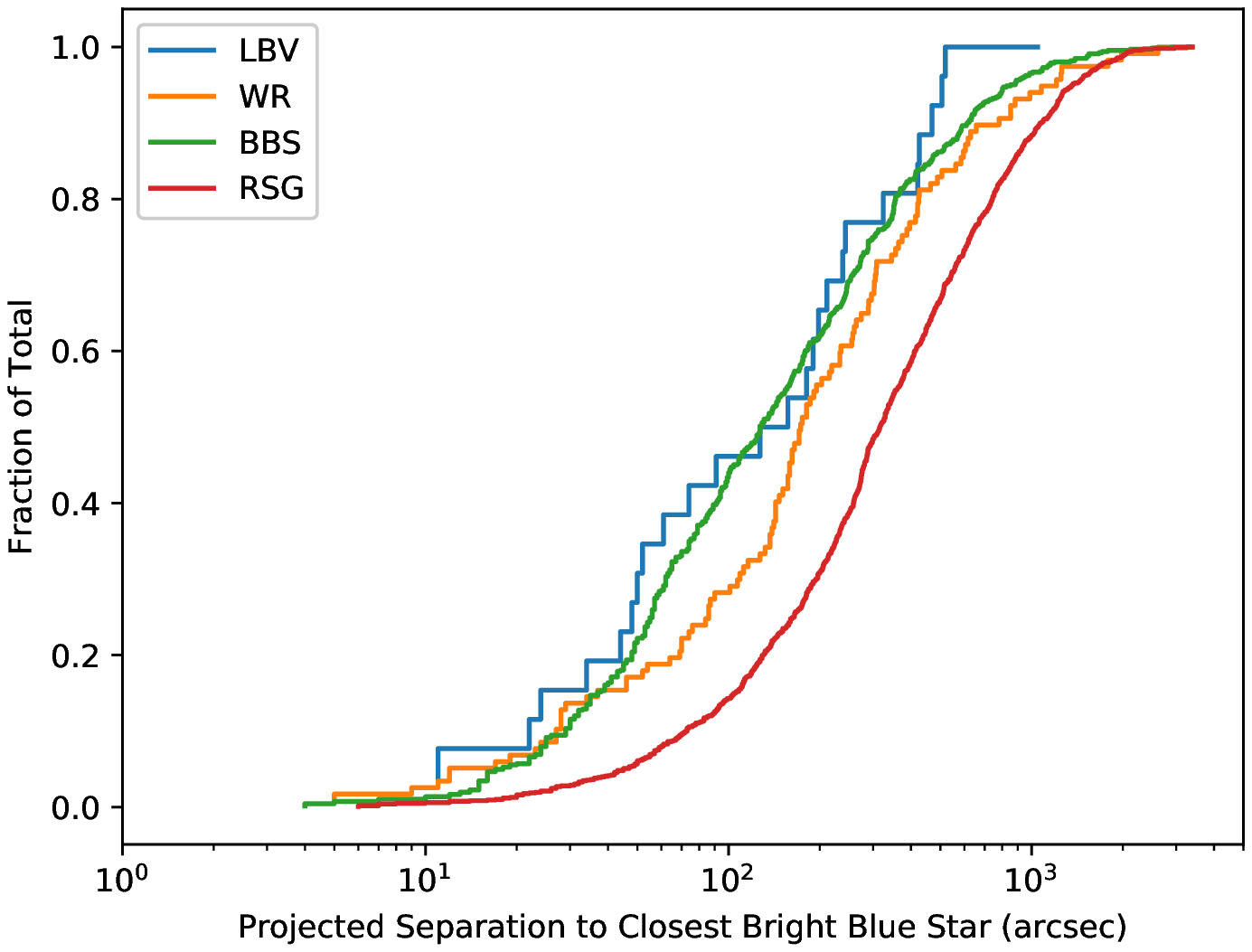}
\caption{\label{fig:LMC}  Projected separations for the stars in the LMC. The fraction of the total number of stars is shown as a function of the projected separation to the nearest neighboring BBS for the LBVs in the LMC, with the projected separation to the nearest neighboring BBS for BBS themselves, WRs, and RSGs for comparison.}
\end{figure}

	As shown in Figure~\ref{fig:LMC}, the projected separation of LBVs and WRs from the nearest BBS are very similar to the projected separation of the BBSs themselves.  In contrast, the RSGs are significantly more isolated. Numerically, the median projected separation for each BBS star from its nearest BBS neighbor is 129\arcsec\ (31 pc).  The LBVs have a median projected separation from the nearest BBS star of 181\arcsec\ (43 pc), and the WRs have a median projected separation from the BBS sample of 175\arcsec\ (42 pc).  By contrast, the separation for the RSGs from the BBS sample is about twice as large, with a median projected separation of 316\arcsec\ (76 pc). 
    
    A KS test yields a p-value of 96\% between the LBV and BBS projected separations, a p-value of 56\% between the LBV and the WR projection separations, while the RSG and the LBV projected separations have a p-value of only 0.3\%.  Since the LBV-RSG KS agreement is well below 5\%, we can reject at high confidence the possibility that the LBV and RSG distributions are from the same parent distribution.  This is consistent with RSGs evolving from lower-mass stars than the LBVs or WRs.

\begin{figure}
\epsscale{1}
\plotone{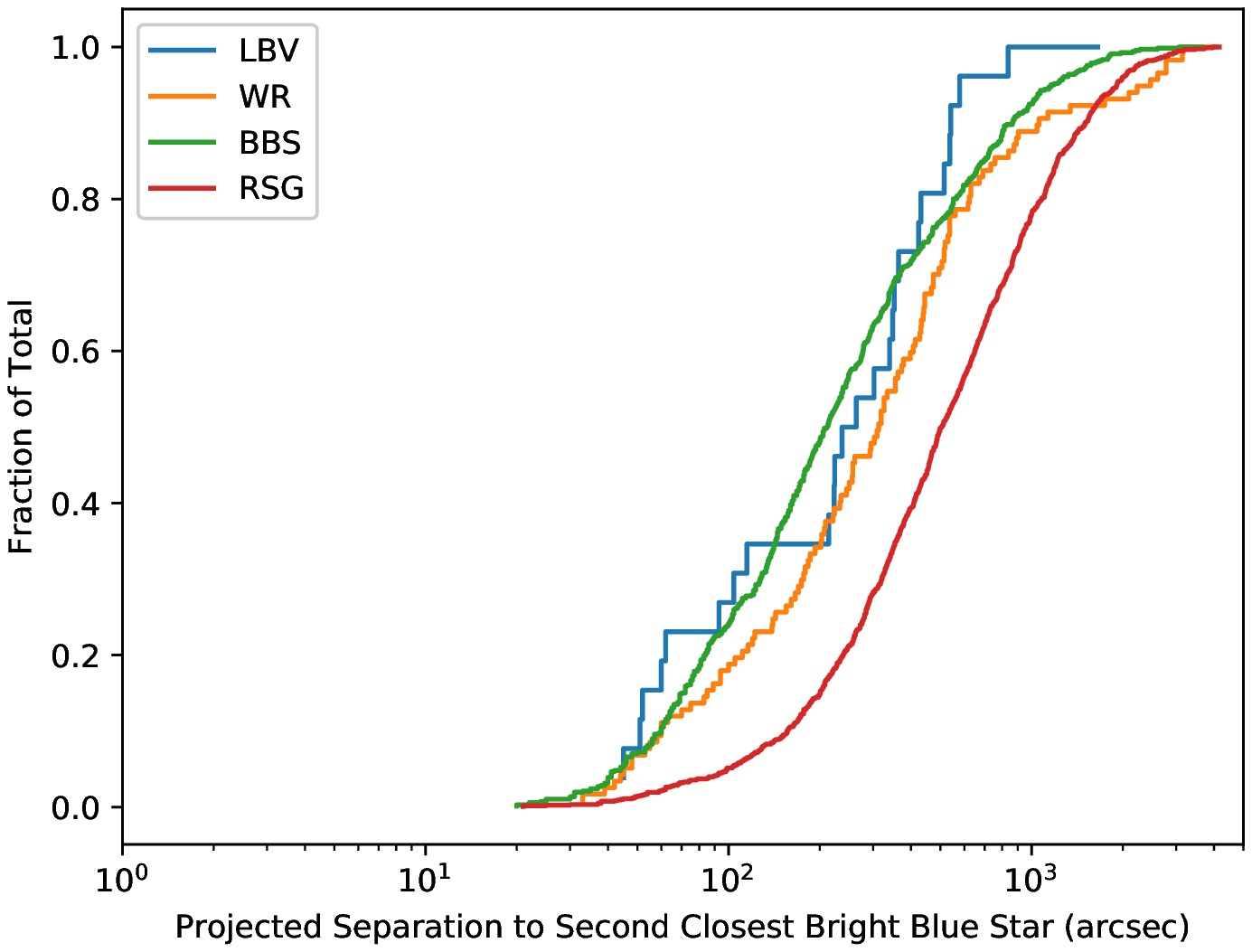}
\caption{\label{fig:LMC_sec}  Projected separations for the stars in the LMC to the second closest BBS. The fraction of the total number of stars is shown as a function of the projected separation to the second nearest neighboring BBS for the LBVs in the LMC, with the projected separation to the second nearest neighboring BBS for BBS themselves, WRs, and RSGs for comparison.}
\end{figure}

 Although this method does not lend itself to a formal analysis, we can at least test the robustness of our results by seeing if the same results hold by using the second-closest neighbor as a measure of isolation. The results are shown in Figure~\ref{fig:LMC_sec}.  We find that the projected separations have similar distributions as before.  The median projection separations of BBSs is 213\arcsec\ (51 pc); for LBVs, it is 302\arcsec\ (72 pc); for WRs, it is 319\arcsec\ (77 pc); and for RSGs, it is again almost double that at 506\arcsec\ (121 pc). The results of a KS test between the LBV distribution and the BBS distribution for the second closest BBSs is 26\%, for WR distribution and the LBVs it is 73\%, and for RSG distribution and the LBVs it is 0.1\%.  This shows that the second closest BBS projected separations have similar results to that of the closest BBS projected separations, and the conclusions hold.

	Another piece of evidence regarding the connection between LBVs and high-mass stars (i.e., the BBS sample) comes from examining their membership in OB associations.  In Table \ref{tab:LMC}, we list the membership status for LBVs in or near the LMC's OB associations \citep{Lucke1970}, using the same criteria as \cite{Neugent2018} did for WR stars\footnote{\cite{Lucke1970} list the size of the associations to the nearest arcminute, as well as giving the 1975 coordinates of the center.  To test for membership, we added one arcminute to the size of the radius and then determined if a star was within 1.2 radii of the precessed coordinates.  If a star was within 2.5 radii, it was considered ``near'' the association and was included.}.  Recall that the OB associations were defined by \cite{Lucke1970} in the usual way (see, e.g., \citealt{Hodge86}), namely based primarily on visual inspection of blue and red pairs of images to identify collections of blue stars (see, e.g., \citealt{HumpSand} for a more complete description of the process).  In other words, the identification of OB associations is not dependent upon the completeness of spectroscopic information. We also calculated the membership rates for our RSG sample and for the BBSs themselves. Excluding stars in 30 Dor, 65\% of LBVs are in or near an OB association, compared to 52\% for BBSs, 68\% for WRs,  and 23\% for RSGs.  Thus, most BBSs, LBVs, and WRs are found in OB associations, while far fewer RSGs are members, as we would expect from their lower masses and greater ages.  We note that a similar conclusion can be drawn from Table 1 of \cite{Walborn2017}, where they find that five of their six active LBVs in the LMC are members of clusters or OB associations. \\

\section{M31/M33}
\label{Sec-M}

\subsection{Sample Selection}
\label{Sec-M-Sample}

	The comparison sample of BBSs was found using the LGGS photometry of M31 and M33 \citep{Massey2007_II,LGGS_sources}.  In order to stay consistent with Section~\ref{Sec-LMC}, the cuts for the sample were kept at $M_V < -5.0$, $(U-B)<-0.5, B-V<0.2$, and $-1.2< Q <-0.88$.  The apparent distance modulus for M31 is 24.8 and for M33 is 25.0 (\citealt{Massey2007_II} and references therein); therefore, the $V$ magnitude cuts became $V<19.8$ and $V<20.0$ for M31 and M33, respectfully.  We expect there will be regions in M31 and M33, like their nuclei, where incompleteness will be an issue, but we expect to find few massive stars there of any kind.
    
	As with the LMC, the LBVs for M31 and M33 were taken from \cite{LBVlist} and are listed in Tables 2 and 3, respectfully.  The RSGs were identified using the LGGS photometry adopting $V-R>0.6$, $V<20$.  As discussed above, this should extend down to a $\log L/L_\odot$ of 4.5, or about 12$M_\odot$.  The WRs were selected based on the classifications from \cite{NeugentM33}, \cite{NeugentM31}, and \cite{LGGS_sources}. 
	    
    Our experience with identifying RSGs in M31 and M33 shows that about half of the red stars in the right magnitude-color range are foreground stars.   As discussed earlier, we have been successful in separating the {\it bona fide} extragalactic RSGs from foreground red dwarfs using a two-color plot.  For cool stars, {\it B-V} becomes primarily a surface gravity indicator due to the effects of line-blanking in the blue, while {\it V-R} remains primarily a temperature indicator, as first discovered by \citet{Massey98} and subsequently utilized in our studies of RSGs of M31 (\citealt{Massey2009, MasseyEvans}) and M33 (\citealt{Drout2012}).   Therefore, we have applied a cut-off of  $B-V > -1.599\times(V-R) + 4.18\times(V-R) - 0.83$ to select supergiants.
    
However, our radial velocities studies have shown that while this method is mostly effective, there are invariably some foreground stars misidentified as RSGs, and some RSGs misidentified as foreground stars.  A quick check of the {\it Gaia} DR2 confirms that a substantial number of our photometrically identified stars are missing in the DR2 as they are too faint.  For the stars with non-negligible parallaxes, most uncertainties are similar to the size of the parallax itself.  We have decided then to include a second sample of RSGs, namely those that have been identified spectroscopically in the LGGS on the basis of their radial velocities.  This is a considerably smaller sample (626 stars), and the distribution does not cover all of M31 as there is the ``alligator jaws'' region in the NE where radial velocities cannot distinguish M31 stars from foreground (\citealt{DroutM31}).  However, the selection is otherwise non-clumpy; i.e., it does not suffer from the same selection effects we expect from the LMC O-type star sample. As we will see that although neither RSG sample is perfect, they yield very similar results.
    
    In M31, there are only six confirmed LBVs (Table~\ref{tab:M31M33}) and in M33, there are only five confirmed LBVs (Table~\ref{tab:M33}).  To avoid small number statistics for the LBVs and to provide a complete picture of the spatial distributions, the LBVs, WRs, BBSs, and RSGs from M31 were combined with their counterparts in M33, resulting in a combined list of the M31 and M33 data for each type of star.  There are 1824 BBSs,  50 LBVs and LBV candidates, 374 WRs, and 4243 photometrically defined RSGs, and 626 spectroscopically confirmed RSGs.

\subsection{Analysis}
\label{Sec-M-Analysis}

\begin{figure}
\epsscale{1}
\plotone{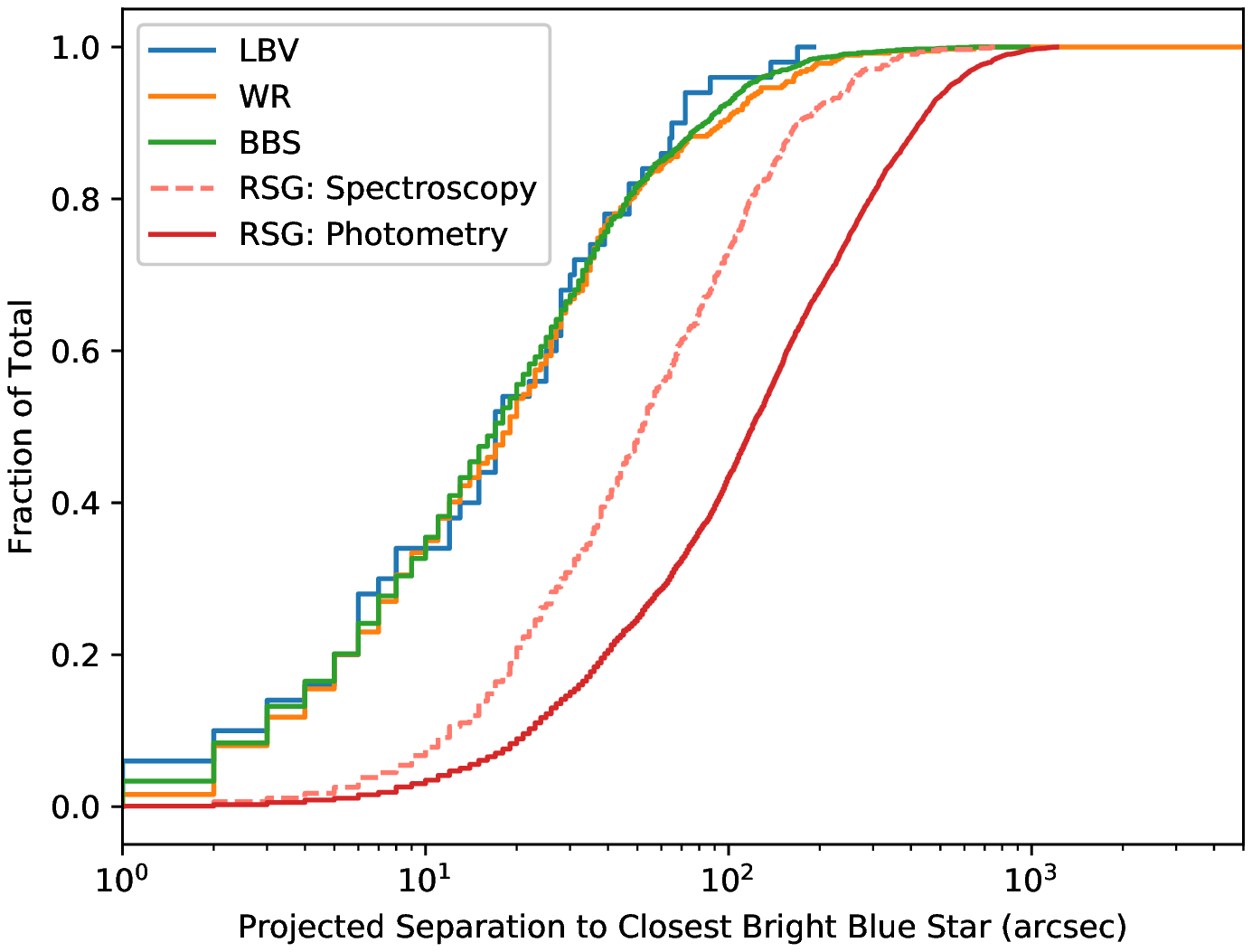}
\caption{\label{fig:M31M33} Projected separations for the stars in M31/M33 to the closest BBS. The fraction of the total number of stars is shown as a function of the projected separation to the nearest neighboring BBS for the LBVs in M31 and M33, with the projected separation to the nearest neighboring BBS for BBS themselves, WRs, and RSGs (both the photometric and spectroscopic defined samples) for comparison.}
\end{figure}

	The projected separations relative to the BBS sample are shown in Figure~\ref{fig:M31M33} for the BBSs, the LBVs, the WRs, and the two RSG samples.  We see immediately that the projected separation of the LBVs from BBSs is similar to that of the BBSs themselves and that of the WRs, while the RSGs have a much larger average separation from the BBSs, as was the case for the LMC.
    
	The median projected separations from BBSs for the BBSs themselves is 17\arcsec\ ($\sim$65 pc).  The LBVs have a median projected separation of 18\arcsec\ ($\sim$68 pc), and the WRs have a median projected separation of 19\arcsec\ ($\sim$72 pc). The photometrically defined RSGs have a median projected separation of 120\arcsec\ ($\sim$456 pc), over five times as large as the other median projected separations.  Even the spectroscopically defined RSGs have a median projected separation of 52\arcsec\ ($\sim$198 pc), twice as large as the other samples.
    
    A KS test shows that the distribution of the LBVs and the BBSs have a p-value of 95\%, the LBVs and the WRs have a p-value of 100\%.  In contrast, the KS test provides p-values well below 0.1\% for both RSG distributions.  We can thus reject the hypothesis that the LBVs and RSGs come from the same parent distribution.  Once again, we find that the LBV isolation is similar to that of the BBSs and WRs,  but unlike that of the RSGs.

\begin{figure}
\epsscale{1}
\plotone{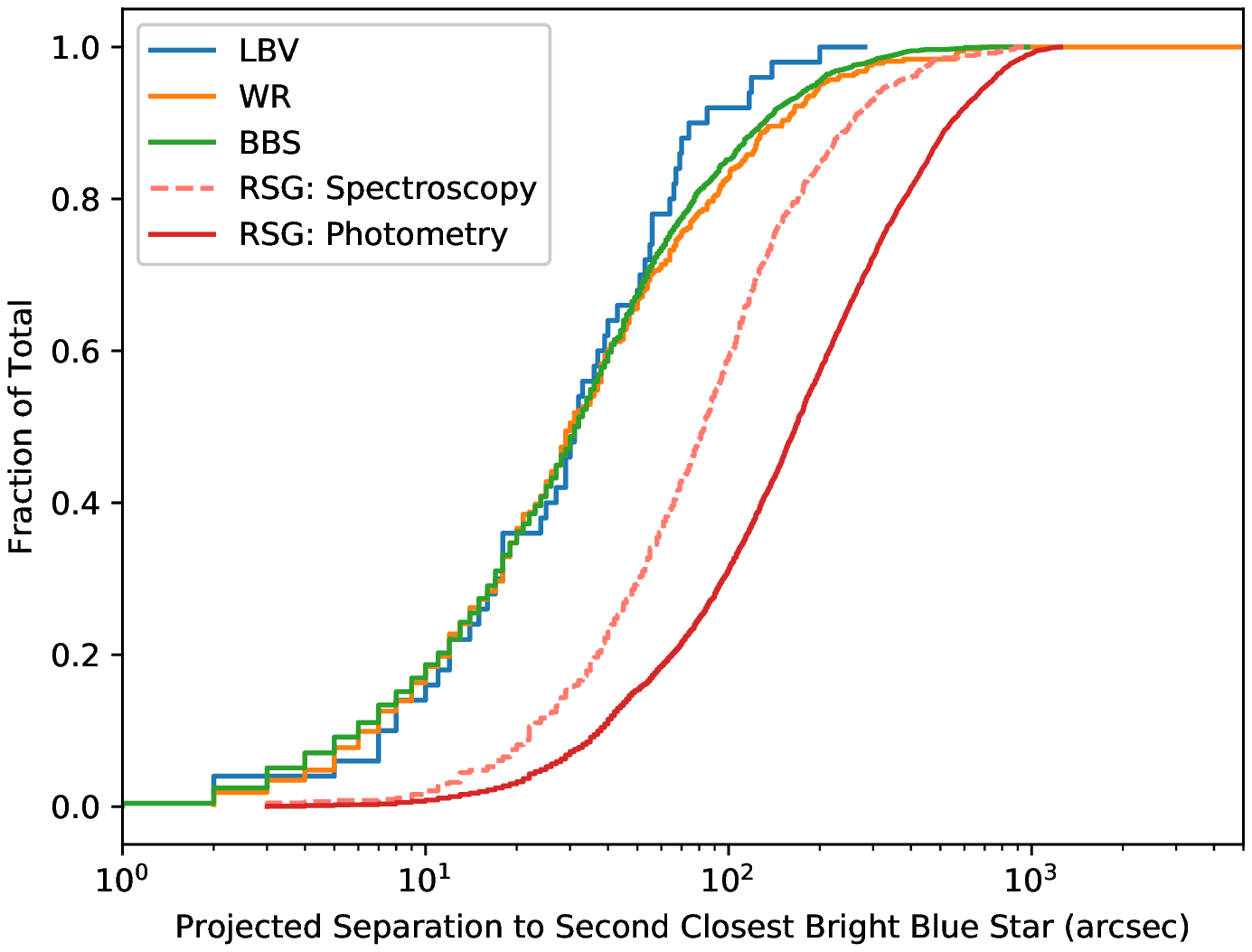}
\caption{\label{fig:M31M33_sec} Projected separations for the stars in M31/M33 to the second closest BBS. The fraction of the total number of stars is shown as a function of the projected separation to the second nearest neighboring BBS for the LBVs in M31 and M33, with the projected separation to the second nearest neighboring BBS for BBS themselves, WRs, and RSGs (both the photometric and spectropic defined samples) for comparison.}
\end{figure}

	The second closest BBS, from the different types of stars was again evaluated and the corresponding projected separations are shown in Figure~\ref{fig:M31M33_sec}.  When we look at the second closest BBS we find that, like in the LMC, the separations have similar relative distributions compared to the closest BBS projected separation.  The median separation of BBSs is 32\arcsec\ (122 pc), that of the LBVs is also 32\arcsec, for WRs it is 30\arcsec\ (114 pc), and for RSG photometric sample it is again around five times that at 167\arcsec\ (635 pc). The spectroscopic sample of RSG has a projected separation of 83\arcsec\ (315 pc). The results of a KS test between the LBVs and the second closest BBSs is 80\%, for WRs it is 52\%, and for RSG it is $<<0.1$\%.  The distributions are once again are similar to the closest projected separation for M31/M33, supporting our original results.

	The M31/M33 results are in agreement with what we found for the LMC, namely that the LBVs show a similar degree of isolation to massive unevolved stars and WRs, and are less isolated than RSGs. \\

\section{Conclusions}
\label{Sec-conclusions}

	The spatial completeness of the \cite{Zaritsky} and LGGS \citep{Massey2007_II,LGGS_sources} catalogs provides a means to create a thorough comparison sample of BBSs.  The photometric sample is able to give a more reliable projected separation measure from the LBVs to the nearest BBSs than the spectroscopic data used in past studies \citep{Smith2015,Humphreys2016}. Overall, this gives us the ability to analyze the spatial distribution of LBVs in a new light\footnote{Further improvements to identifying the most massive unevolved stars are being made possible by such wide-field, high resolution studies as the Panchromatic Hubble Andromeda Treasury survey of M31 \citep{PHAT}.}.

    The analysis for both the LMC and M31/M33 reveals that the LBVs are not isolated relative to their expected progenitors, high-mass unevolved stars, or to their expected descendants, the WRs. By contrast, the RSGs are considerably more isolated, as one would expect in the standard evolutionary picture where they come from stars of lower mass than do the stars that make up the BBS, LBV, and WR samples. 

	We have not included either the SMC or the Milky Way LBVs in this study.   The Small Magellanic Cloud (SMC) contains only two confirmed LBVs.  Still, we note that HD 5980 (one of the LBVs in the SMC) is located in the outskirts of NGC 346, a cluster rich in very early massive stars; see \cite{Massey1989}\footnote{\cite{Smith2015} argues that HD~5980 is not truly a member of NGC 346, but is well isolated from O-type stars in the cluster.  We refer the reader to  Figure~1 in \cite{Massey1989}, where HD 5980 is identified as AV 229.  Although not in the central region, it is certainly well embedded in the cluster's nebulosity, and is no further away from the core than other prominent O-type stars, such as AV 232 (Sk 80), an O7~Iaf+ \citep{Walborn1977} which is the cluster's second most luminous and massive star (after HD 5980; see Tables VI and XI of \citealt{Massey1989}), and is coeval with the rest of the cluster. That said, we note that HD~5980 is a very complex system, with multiple binary components (see, e.g., \citealt{Ko2014}), and may not serve as a good archetype for the LBV phenomenon.}.  The availability of distances for Galactic stars thanks to {\it Gaia} presents an opportunity for exploring this further for Milky Way LBVs; \cite{Smith2018Gaia} has made an important step in investigating the distances to Galactic LBVs.  These are found out to a distance of 7~kpc;  a great deal of work needs to be done in identifying a suitable reference sample of unevolved massive stars within this volume.  
   
\newpage
    \cite{Smith2015} state that it was unlikely that all LBVs are a transitionary phase between O-type stars and WR stars, because it would require dispersing and then un-dispersing.  However, our results suggest that LBVs have not dispersed far from their place of origin, which supports the standard picture that LBVs are a shorted-lived stage in the lives of massive stars.  Of course, LBVs could still be mass gainers in binary systems, but our results show that LBVs are less dispersed than RSGs, arguing for either higher progenitor masses and/or significantly younger ages. The majority of LBVs are also contained in OB associations (likely birthplaces), which further suggests that they have not traveled far in their evolutionary lifetimes.  Any proposed binary formation channel would need to also satisfy these constraints.
    
That said, we do have concerns about the use of spatial separations as a means of uncovering the evolutionary connection between objects.  At the very least, the selection of the various sets of stars need to be complete.  Other than in areas of extreme crowding, the BBS populations of the LMC, M31, and M33 should be complete. Similarly our knowledge of the WR populations of the LMC, M31, and M33 are known at the 95\% (or better) level.  However, the identifications of LBVs and LBV candidates in these galaxies strike us as quite uncertain.  True LBVs identify themselves by spectacular outbursts, but these may occur only once every few centuries.   LBV candidates are selected on the basis of spectroscopic similarity to sanctified LBVs, but only a tiny fraction of stars in the appropriate magnitude/color range have been examined spectroscopically in any of these galaxies.  Our understanding of the LBV phenomenon is constantly evolving: who expected an Ofpe/WN9 star to become an LBV before R127 had its major outburst in 1980 (\citealt{Walborn2017} and references therein)?  We do not know what other types of stars may surprise us in this same way.

\acknowledgments
We are very grateful to Dr.\ Emily Levesque for offering comments and advice on the paper.
This work was supported by the National Science Foundation through AST-1612874 and support was provided to MRD by NASA through Hubble Fellowship grant NSG-HF2-51373 awarded by the Space Telescope Science Institute, which is operated by the Association of Universities for Research in Astronomy, Inc., for NASA, under contract NAS5-26555. MRD also acknowledges support from the Dunlap Institute at the University of Toronto.  This work makes use of data products from the Two Micron All Sky Survey, which is a joint project of the University of Massachusetts and the Infrared Processing and Analysis Center/California Institute of Technology, funded by the National Aeronautics and Space Administration and the National Science Foundation. It has also made use of data from the European Space Agency (ESA) mission {\it Gaia} (\url{https://www.cosmos.esa.int/gaia}), processed by the {\it Gaia} Data Processing and Analysis Consortium (DPAC, \url{https://www.cosmos.esa.int/web/gaia/dpac/consortium}). Funding for the DPAC has been provided by national institutions, in particular the institutions participating in the {\it Gaia} Multilateral Agreement.

\clearpage

\begin{deluxetable}{l l c c c c c c c}
\tablecaption{\label{tab:LMC} LBVs in the LMC}
\tablewidth{0pt}
\tablehead{
\colhead{Star Name}
&\colhead{Classification}
&\colhead{R.A.}
&\colhead{Decl.}
&\colhead{\textit{V} mag$^a$}
&\colhead{OB Assoc.$^b$}
&\colhead{Humphreys$^c$}
&\colhead{Smith$^d$}
&\colhead{Richardson$^a$}
}
\startdata
S Dor  & LBV &  05:18:14.36 & -69:15:01.1 & 10.25 & 41 & \checkmark & \checkmark & \checkmark \\
RMC 127   & LBV &  05:36:43.69 &  -69:29:47.4 & 10.15 & 94 & \checkmark & \checkmark & \checkmark \\
RMC 143   & LBV &  05:38:51.62 &  -69:08:07.3 & 12.01 & 100 & \checkmark & \checkmark & \checkmark \\
RMC 71    & LBV &  05:02:07.39 &  -71:20:13.1 &  10.55 & \nodata & \checkmark & \checkmark & \checkmark \\
RMC 110   & LBV &  05:30:51.48 &  -69:02:58.6 & 10.28 & \nodata & \checkmark & \checkmark & \checkmark \\
RMC 85    & candidate &  05:17:56.07 &  -69:16:03.8 & 10.84 & 41 & \checkmark & \checkmark & \checkmark \\
Sk -67$^\circ$266 & candidate & 05:45:51.94 & -67:14:25.9 & 11.95 & 116 & \checkmark & \checkmark & \checkmark \\
HD 269687 & candidate & 05:31:25.52 & -69:05:38.6 & 11.87 & \nodata & \checkmark & \checkmark & \checkmark \\
Sk -69$^\circ$142a  & LBV & 05:27:52.66 &  -68:59:08.5 & 10.73 & \nodata & \checkmark & \checkmark & \checkmark \\
Sk -69$^\circ$279  & candidate & 05:41:44.66 &  -69:35:14.9 & 12.84 & \nodata & \checkmark & \checkmark&\checkmark\\
RMC 81  & candidate & 05:10:22.79 &  -68:46:23.8 & 10.52 & \nodata & \nodata & \checkmark & \checkmark\\
RMC 84       & candidate & 05:13:54.28 &  -69:31:46.7 & 12.71 & 39 & \nodata & \checkmark & \checkmark \\
RMC 99       & candidate & 05:22:59.79 &  -68:01:46.6 & 11.51 & 49 & \nodata & \checkmark & \checkmark\\
RMC 126      & candidate & 05:36:25.85 &  -69:22:55.8 & 10.99 & 93 & \nodata & \checkmark & \checkmark\\
Sk -69$^\circ$271  & candidate & 05:41:20.13 &  -69:36:22.9 & 11.79 & (103) & \nodata & \checkmark & \checkmark\\
RMC 66       & candidate & 04:56:47.08 &  -69:50:24.8 & 10.63 & \nodata & \nodata & \nodata & \checkmark\\
RMC 74       & candidate & 05:04:14.91 &  -67:15:05.2 &  11.03 & 19 & \nodata & \nodata & \checkmark\\
Sk -68$^\circ$42   & candidate & 05:05:53.98 &  -68:10:50.5 & 12.07 & (25) & \nodata & \nodata & \checkmark\\
RMC 78       & candidate & 05:07:20.42 &  -68:32:08.6 & 11.54 & \nodata & \nodata & \nodata & \checkmark\\
HD 269216    & LBV & 05:13:30.78 &  -69:32:23.6 & 11.12 & 39 & \nodata & \nodata & \checkmark\\
MWC 105    & candidate & 05:13:52.99 &  -67:26:54.8 & 11.59 & 38 & \nodata & \nodata & \checkmark\\
Sk -68$^\circ$93   & candidate & 05:28:31.37 &  -68:53:55.7 & 10.74 & (64) & \nodata & \nodata & \checkmark\\
RMC 116$^e$     & candidate & 05:31:52.28 &  -68:32:38.9 & 10.54 & \nodata & \nodata & \nodata & \checkmark\\
RMC 123      & candidate & 05:35:16.63 &  -69:40:38.4 & 10.69  & 87 & \nodata & \nodata & \checkmark\\
RMC 128      & candidate & 05:36:47.19 &  -69:29:52.1 & 10.73 & 94 & \nodata & \nodata & \checkmark\\
RMC 149      & candidate & 05:39:58.75 &  -69:44:04.1 &  12.52 & 105 & \nodata & \nodata & \checkmark\\
MWC 126    & candidate & 05:40:13.32 &  -69:22:46.5 & 11.93 & 104 & \nodata & \nodata & \checkmark\\
MWC 112     & LBV & 05:28:21.97 & -68:59:48.3 & 11.45 & \nodata & \nodata & \checkmark & \nodata \\
\enddata
\tablenotetext{a}{From \cite{SIMBAD}}
\tablenotetext{b}{Lucke-Hodge (LH) OB association numbers are from \cite{Lucke1970}.  Parenthesis are used to denote the association if the star is only near the association. Note that the 30 Dor region corresponds to LH 100 and its extension to the SW LH 99, while the center of Constellation III is LH 84.}
\tablenotetext{c}{From \cite{Humphreys2016}}
\tablenotetext{d}{From \cite{Smith2015}}
\tablenotetext{e}{\cite{LBVlist} has RMC 116 listed as an LBV, here we have it listed as a candidate due to the lack of evidence supporting the LBV status.}

\end{deluxetable}

\begin{deluxetable}{l l c c c c}
\tablecaption{\label{tab:M31M33} LBVs in M31}
\tablewidth{0pt}
\tablehead{
\colhead{Star Name}
&\colhead{Classification}
&\colhead{R.A.}
&\colhead{Decl.}
&\colhead{\textit{V} mag$^a$}
&\colhead{Alternate Name}
}
\startdata
AE And   & LBV &  00:43:02.51 & +41:49:12.2 & 17.43 & \nodata \\
AF And   & LBV &  00:43:33.08 & +41:12:10.3 & 17.33 & \nodata \\
Var 15   & LBV &  00:44:19.42 & +41:22:47.0 & 18.45 & \nodata \\
J004526.62+415006.3 & LBV & 00:45:26.61 & +41:50:06.1 & 16.39 & UCAC4 660-003111 \\
Var A-1 & LBV & 00:44:50.53 & +41:30:37.6 & 17.14 & \nodata \\
J003910.85+403622.4 & candidate & 00:39:10.84 & +40:36:22.3 & 18.18 & \nodata \\
J004051.59+403303.0 & LBV  & 00:40:51.58 & +40:33:02.9 & 16.99 & \nodata \\
J004322.50+413940.9 & candidate & 00:43:22.49 & +41:39:40.8 & 20.35 & \nodata \\
J004341.84+411112.0 & candidate  & 00:43:41.83 & +41:11:11.9 & 17.55 & \nodata \\
J004350.50+414611.4 & candidate & 00:43:50.49 & +41:46:11.2 & 17.74 & \nodata \\
J004411.36+413257.2 & candidate & 00:44:11.35 & +41:32:57.1 & 18.07 & {[}MLV92{]} 339869 \\
J004425.18+413452.2 & candidate & 00:44:25.17 & +41:34:52.1 & 17.48 & {[}WB92a{]} 411 \\
J004444.01+415152.0 & candidate & 00:44:44.00 & +41:51:51.8 & 19.03 & \nodata \\
J004507.65+413740.8 & candidate & 00:45:07.64 & +41:37:40.7 & 16.15 & \nodata \\
J004522.58+415034.8 & candidate & 00:45:22.57 & +41:50:34.6 & 18.50 & \nodata \\
\enddata
\tablenotetext{a}{From \cite{LGGS_sources}}
\end{deluxetable}

\begin{deluxetable}{l l c c c c}
\tablecaption{\label{tab:M33} LBVs in M33}
\tablewidth{0pt}
\tablehead{
\colhead{Star Name}
&\colhead{Classification}
&\colhead{R.A.}
&\colhead{Decl.}
&\colhead{\textit{V} mag$^a$}
&\colhead{Alternate Name}
}
\startdata
Var B  & LBV &   01:33:49.20  & +30:38:09.0 & 16.21 & \nodata \\
Var C  & LBV &   01:33:35.11  & +30:36:00.3 & 16.43 & \nodata \\
Var 83 & LBV &   01:34:10.90  & +30:34:37.5 & 16.03 & \nodata \\
Var 2  & LBV &   01:34:18.34  & +30:38:36.9 & 18.22 & \nodata \\
J013229.03+302819.6 & candidate & 01:32:29.00 & +30:28:19.5 & 19.00 & \nodata \\
J013235.25+303017.6 & candidate & 01:32:35.22 & +30:30:17.5 & 18.01 & \nodata \\
J013245.41+303858.3 & candidate & 01:32:45.38 & +30:38:58.2 & 17.61 & UIT 008 \\
J013248.26+303950.4 & candidate & 01:32:48.23 & +30:39:50.3 & 17.25 & \nodata \\
J013300.02+303332.4 & candidate & 01:32:59.99 & +30:33:32.3 & 18.32 & {[}HS80{]} B43 \\
J013303.09+303101.8 & candidate & 01:33:03.06 & +30:31:01.7 & 16.99 & {[}HS80{]} B48 \\
J013317.05+305329.9 & candidate & 01:33:17.02 & +30:53:29.8 & 18.92 & FSZ 83 \\
J013317.22+303201.6 & candidate & 01:33:17.19 & +30:32:01.5 & 18.75 & \nodata \\
J013332.64+304127.2 & candidate & 01:33:32.61 & +30:41:27.1 & 18.99 & {[}DMS93{]} NGC 595 6 \\
J013334.06+304744.3 & candidate & 01:33:34.03 & +30:47:44.2 & 17.46 & FSZ 126 \\
J013337.35+303329.0 & candidate & 01:33:37.32 & +30:33:28.9 & 18.51 & \nodata \\
J013339.52+304540.5 & candidate & 01:33:39.49 & +30:45:40.4 & 17.50 & FSZ163 \\
J013341.28+302237.2 & candidate & 01:33:41.25 & +30:22:37.1 & 16.29 & {[}HS80{]} 110A  \\
J013344.79+304432.4 & candidate & 01:33:44.76 & +30:44:32.3 & 18.15 & {[}MAP95{]} M33 OB66 28 \\
J013351.46+304057.0 & candidate & 01:33:51.43 & +30:40:56.9 & 17.73 & IFM-B 1079 \\
J013352.42+303909.6 & candidate & 01:33:52.39 & +30:39:09.5 & 16.17 & {[}MJ98{]} WR 98 \\
J013354.85+303222.8 & candidate & 01:33:54.82 & +30:32:22.7 & 18.34 & {[}MJ98{]} WR 98 \\
J013355.96+304530.6 & candidate & 01:33:55.93 & +30:45:30.5 & 14.86 & {[}HS80{]} B324 \\
J013357.73+301714.2 & candidate & 01:33:57.70 & +30:17:14.1 & 17.39 & \nodata \\
J013406.63+304147.8 & candidate & 01:34:06.60 & +30:41:47.7 & 16.08 & {[}HS80{]} B416 \\
J013406.80+304727.0 & candidate & 01:34:06.77 & +30:47:26.9 & 17.20 & {[}HS80{]} B393 \\
J013406.80+304727.0 & candidate & 01:34:16.04 & +30:36:42.0 & 17.95 & {[}HS80{]} B517  \\
J013416.10+303344.9 & candidate & 01:34:16.07 & +30:33:44.8 & 17.12 & {[}HS80{]} B526 \\
J013416.44+303120.8 & candidate & 01:34:16.41 & +30:31:20.7 & 17.10 & \nodata \\
J013422.91+304411.0 & candidate & 01:34:22.88 & +30:44:10.9 & 17.22 & FSZ 458 \\
J013424.78+303306.6 & candidate & 01:34:24.75 & +30:33:06.5 & 16.84 & Pul -3 120290 \\
J013427.26+304600.1 & candidate & 01:34:27.23 & +30:46:00.0 & 19.24 & FSZ 465 \\
J013429.64+303732.1 & candidate & 01:34:29.61 & +30:37:32.0 & 17.11 & \nodata \\
J013432.76+304717.2 & candidate & 01:34:32.73 & +30:47:17.1 & 19.09 & \nodata \\
J013459.39+304201.2  & candidate & 01:34:59.36 & +30:42:01.1 & 18.25 & {[}MAP95{]} M33 OB88 7 \\
J013509.73+304157.3 & LBV &  01:35:09.70  & +30:41:57.2 & 18.04 & M33 V0532 \\
\enddata
\tablenotetext{a}{From \cite{LGGS_sources}}
\end{deluxetable}

\clearpage
\bibliographystyle{apj}
\bibliography{LBVPaper}

\end{document}